\documentclass[11pt,a4paper]{article} 
\usepackage{epsfig} %include package for eps-files

\title{The Interaction of Two Hopf Solitons}  \author{R S Ward,
\\Department of Mathematical Sciences,  \\ University of
Durham, \\Durham DH1 3LE \\email: richard.ward@durham.ac.uk} 

  \hoffset =
-2cm \textwidth = 170mm

\begin{document}
\maketitle \abstract{This Letter deals with topological solitons in an
O(3) sigma model in three space dimensions (with a Skyrme term to
stabilize their size).  The solitons are classified topologically by their
Hopf number $N$.  The $N=2$ sector is studied; in particular, for two
solitons far apart, there are three ``attractive channels''.
Viewing the solitons as dipole pairs enables one to predict the force
between them.  Relaxing in the attractive channels leads to various static
2-soliton solutions.
}

\section{Introduction}

The O(3) nonlinear sigma model in 3+1 dimensions involves a unit vector
field $\vec{\phi} = (\phi^1,\phi^2,\phi^3)$ which is a
function of the space-time coordinates $x^\mu = (t,x,y,z)$.  Since $\vec{\phi}$
takes values on the unit sphere $S^2$ and the homotopy group
$\pi_3(S^2)$ is non-trivial, the system admits topological
solitons (textures) with non-zero Hopf number $N \in \pi_3(S^2)$.
The model arises naturally in condensed-matter physics and in cosmology,
and its dynamics in these contexts is governed by the Lagrangian density
$(\partial_\mu \vec{\phi})^2$.  With such a Lagrangian, there are no stable
soliton solutions: the usual scaling argument shows that
textures are unstable to shrinking in size.  (In practice, the decay of these
textures is more complicated, and can, for example, involve decay into
monopole-antimonopole pairs.)

Another context in which the  nonlinear sigma model arises is as an
approximation to a gauge theory; {\it ie} one gets an effective action
for the gauge field in terms of  the scalar fields $\phi^a$.
In this case, there are higher-order terms in the Lagrangian, which can have
the effect of stabilizing solitons.  This idea is familiar in the Skyrme
model, where the target space is a Lie group; it also occurs for the case
relevant here, where the target space is $S^2$  \cite{fn:99}, \cite{s:99},
\cite{clp:99}.

The simplest such modification involves adding a fourth-order
(Skyrme-like) term to the Lagrangian.  This leads to a sigma model which
admits stable, static, localized solitons --- they resemble closed strings,
which can be linked or knotted.  The system has been written about at least
since 1975  \cite{f:75},  \cite{v:78}; but recently interest in it has
increased, stimulated by numerical work \cite{fn:97}, \cite{gh:97},
\cite{bs:97}, \cite{bs:98a}, \cite{bs:98b}, \cite{hs:99}; see also \cite{w:99},
\cite{g:98}, \cite{n:99}.

In this Letter, we shall deal only with static configurations, so
$\vec{\phi}$ is a function of the spatial coordinates  $x^j = (x,y,z)$.
The boundary condition is $\phi^a\to(0,0,1)$ as $r\to\infty$, where
$r^2 = x^2+y^2+z^2$.  So we may think of $\vec{\phi}$ as a smooth function
from $S^3$ to $S^2$; and hence it defines a Hopf number $N$ (an integer).
This $N$ may be thought of as a linking number: the inverse images of two
generic
points on the target space, for example $(0,0,1)$ and $(0,0,-1)$, are curves
in space, and the first curve links $N$ times around the other.

The energy of a static field $\vec{\phi}(x^j)$ is taken to be
\begin{equation} \label{energy}
 E = \frac{1}{32\pi^2}
       \int \bigl[ (\partial_j\phi^a)(\partial_j\phi^a)
         + F_{jk} F_{jk} \bigr] \, d^3x,
\end{equation}
where $F_{jk} = \varepsilon_{abc} \phi^a 
         (\partial_j \phi^b) (\partial_k \phi^c)/2$.
The ratio of the coefficients of the two terms in (\ref{energy}) sets the
length scale --- in this case, one expects the solitons to have a size of
order unity (note that other authors use slightly different coefficients).
The factor of $1/32\pi^2$ is justified in \cite{w:99}: 
there is a lower bound on the energy which is proportional to $N^{3/4}$,
and if space is allowed to be a three-sphere, then there is an $N=1$ solution
with $E=1$.  So one expects, with the normalization (\ref{energy}),
to have the lower bound $E\geq N^{3/4}$.

\section{The One-Soliton}

The minimum-energy configuration in the $N=1$ sector is an axially-symmetric,
ring-like structure.
It was studied numerically in \cite{fn:97} (with no quantitative results),
in \cite{gh:97} (which gave an energy value of $E=1.25$, although without any
statement of numerical errors), and in \cite{bs:98a}, \cite{bs:98b} (where
the field was placed in a finite-volume box, so its energy was not
evaluated accurately).  For the results described in this letter, a
numerical scheme has been set up which
\begin{itemize}
\item includes the whole of space $R^3$
(by making coordinate transformations that bring spatial infinity in to
a finite range); and 
\item using a lattice expression for the energy in which the truncation
error is of order $h^4$, where $h$ is the lattice spacing.
\end{itemize}
Using this shows that the energy of the one-soliton is
$E = 1.22$ (accurate to the two decimal places).

Let us choose the axis of symmetry to be the $z$-axis, and the
soliton to be concentrated in the $xy$-plane.  In terms of the complex field
\begin{equation} 
 W = \frac{\phi^1+i\phi^2}{1+\phi^3}
\end{equation} 
(the stereographic projection of $\vec{\phi}$), the 1-soliton solution is
closely approximated by the expression
\begin{equation}\label{1-sol}
 W = \frac{x+iy}{z-if(r)},
\end{equation} 
where $f(r)$ is a cubic polynomial.  We may minimize the energy of the
configuration (\ref{1-sol}) with respect to the coefficients of $f$: this
gives an energy $E=1.23$ (less than 1\% above the true minimum), for
\begin{equation} 
 f(r) = 0.453(r-0.878)(r^2 + 0.705r + 1.415).
\end{equation}
Note that $W\to 0$ as $r\to\infty$ (the boundary condition); that $W=0$
on the $z$-axis; and that $W=\infty$ on a ring (of radius 0.878) in the
$xy$-plane.  The ring where $W=\infty$ links once around the ``ring''
(the  $z$-axis plus a point at infinity) where  $W=0$: hence the linking
number $N$ equals 1.  The field looks like a ring (or possibly a disc,
depending on what one plots) in the $xy$-plane.

To leading order as $r\to\infty$, $\phi^1$ and $\phi^2$ have to be
solutions of the Laplace equation  ($\phi^1$ and $\phi^2$ are the analogues
of the massless pion felds in the Skyrme model).  From (\ref{1-sol}) one
sees that
\begin{equation} 
 \phi^1+i\phi^2 \approx 2W \approx \frac{4i(x+iy)}{r^3} 
          \quad{\rm for\ large\ }r.
\end{equation}
So  $\phi^1$ and $\phi^2$ resemble, asymptotically, a pair $(\vec{P},\vec{Q})$
of dipoles, orthogonal to each other and to the axis of symmetry.

It is useful to note
the effect on the 1-soliton field of rotations by $\pi$ about each
of the three coordinate axes.  These, together with the identity, form the
dihedral group $D_2$.  Let $\pm I$ and $\pm C$ denote the maps
\begin{equation} 
 \pm I : W \mapsto \pm W, \qquad \pm C : W \mapsto \pm\overline{W}.
\end{equation}
Then it is clear from (\ref{1-sol}) that
the four elements of $D_2$ induce the four maps $\{I,-I,C,-C\}$ on $W$.

The single soliton depends on six parameters: three for location in space,
two for the direction of the $z$-axis, and one for a phase (the phase of $W$).
The phase is unobservable, in the sense that it does not appear in the energy
density, and can be removed by a rotation of the target space $S^2$; but the
energy of a two-soliton system depends on the relative phase of the two
solitons, as we shall see in the next section.

%%%%%%%%%%%%%%%%%%%%%%%%%%%%%%%%%%%%%%%%%%%%
\section{Two Solitons Far Apart}

Suppose we have two solitons, located far apart.  Let $(\vec{P}_+,\vec{Q}_+)$
denote the dipole pair of one of them, $(\vec{P}_-,\vec{Q}_-)$
the dipole pair of the other, and  $\vec{R}$ the separation vector between
them. There will, in general, be a force between the solitons, which depends
(to leading order) on the distance $R$ between them, and on their mutual
orientation.
One can predict this force by considering the forces between the dipoles.
Since the fields are space-time scalars, like charges attract; so the force
between two dipoles is maximally attractive if they are parallel, and
maximally repulsive if they are anti-parallel.
There are three obvious ``attractive channels'' (mutual orientations for which
the two solitons attract), which will be referred to as channels $A$, $B$
and $C$.  We now discuss each of these.

\paragraph{Channel A.}
The only axisymmetric configuration involving two separated solitons
is one where each dipole pair is
orthogonal to the separation vector: in fact, where
$\vec{P}_+\times\vec{Q}_+$, $\vec{P}_-\times\vec{Q}_-$ and $\vec{R}$ are
all parallel.  The configuration is illustrated in Fig 1.
\begin{figure}
\caption{Two solitons in channel A, and the corresponding static 2-soliton
         solution.}
\label{fig1}
\begin{center}
\includegraphics[angle=-90,scale=1,width=0.5\textwidth]{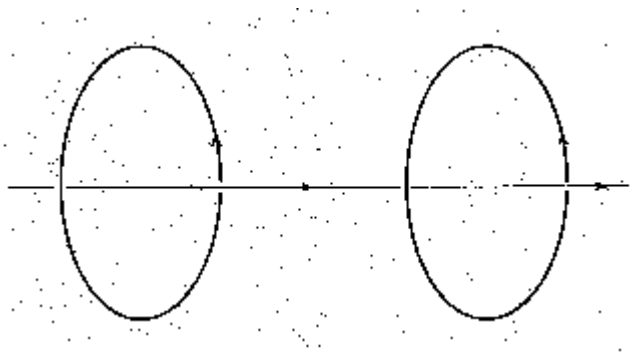}%{fig1.epsi}
\end{center}
\end{figure}
The two circles are where $W=\infty$, while the line linking them is where
$W=0$.  The arrows on the curves serve partly to distinguish solitons from
anti-solitons: the convention here is that solitons obey the right-hand rule,
whereas anti-solitons would obey the left-hand rule.

Let $\theta$ denote the angle between $\vec{P}_+$ and $\vec{P}_-$ ({\it ie}
the pair $(\vec{P}_+,\vec{Q}_+)$ is rotated by $\theta$ about the line
joining the two solitons).  Let $E_1$ denote the energy of a single soliton,
and $E_2(R,\theta)$ the energy of the two-soliton system, as a function
of the separation $R$ and the relative phase $\theta$.  Considering the
potential energy of the interacting dipoles suggests that
\begin{equation}\label{e2A}
 E_2(R,\theta) = 2E_1 - 2k R^{-3} \cos\theta
\end{equation}
for some constant $k$.  Clearly $E_2(R,\theta)$ is minimized, for a given
$R$, when $\theta=0$  ({\it ie} when the two solitons are in phase): this is
channel $A$.
The formula (\ref{e2A}) was tested numerically, by computing the energy
of the configurations obtained by combining translated and rotated versions of
the approximate one-soliton  (\ref{1-sol}).  The combination anstaz was simply
that of addition ($W = W_+ + W_-$), which is a plausible approximation for
large
$R$ (bearing in mind that the $W$-field tends to zero away from each soliton).
For $R$ in the range $6<R<16$, the form (\ref{e2A}) is indeed found to hold,
with $k\approx 1$. (In view of the crudity of the ``sum'' ansatz, the accuracy
is not claimed to be better than 10\% or so; but the  $R^{-3} \cos\theta$
dependence is very clear.)  The behaviour under the discrete symmetries $D_2$
is the same as for the 1-soliton, namely  $\{I,-I,C,-C\}$.

\paragraph{Channel B.}
This channel is one in which both dipole pairs are co-planar with $\vec{R}$,
with $\vec{P}_+\times\vec{Q}_+$ and $\vec{P}_-\times\vec{Q}_-$ being parallel
(and orthogonal to  $\vec{R}$).  This configuration is depicted in Fig 2(a).
\begin{figure}
\caption{Two solitons in attractive channel B coalesce to form a static
     2-soliton solution.  The pictures (a), (b) and (c) are a time-sequence.}
\label{fig2}
\begin{center}
\includegraphics[angle=-90,scale=1,width=0.5\textwidth]{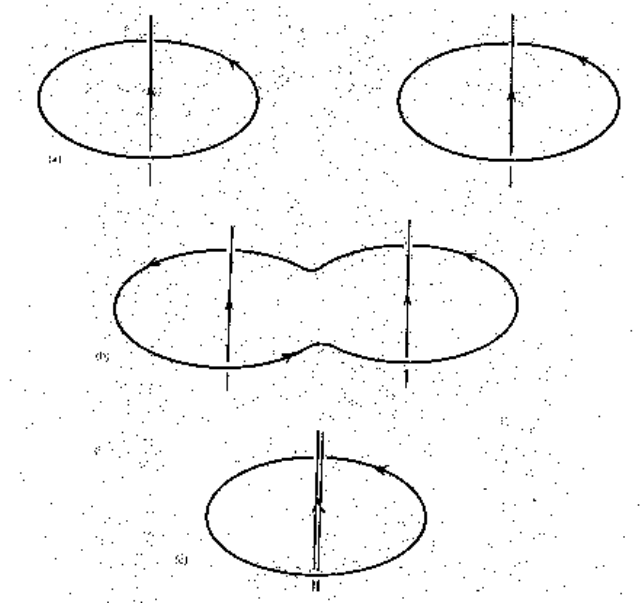}
\end{center}
\end{figure}
In this case, the effect of the discrete symmetries $D_2$ on the configuration
is  $\{I,I,C,C\}$.  Consideration
of the forces between the dipoles suggests that the energy behaves like
\begin{equation}\label{e2B}
 E_2(R,\theta) = 2E_1 + k R^{-3} \cos(\theta_+ - \theta_-), 
\end{equation}
where $\theta_{\pm}$ is the angle that $\vec{P}_{\pm}$ makes
with $\vec{R}$.  The expression (\ref{e2B}) has a minimum when
$\theta_+ -\theta_- = \pi$, and this is attractive channel $B$.

\paragraph{Channel C.}
Here the dipole pairs are again co-planar with $\vec{R}$, but now
$\vec{P}_+\times\vec{Q}_+$ and $\vec{P}_-\times\vec{Q}_-$ are anti-parallel.
This is depicted in Fig 3(a).
\begin{figure}
\caption{Two solitons in attractive channel C.}
\label{fig3}
\begin{center}
\includegraphics[angle=-90,scale=1,width=0.5\textwidth]{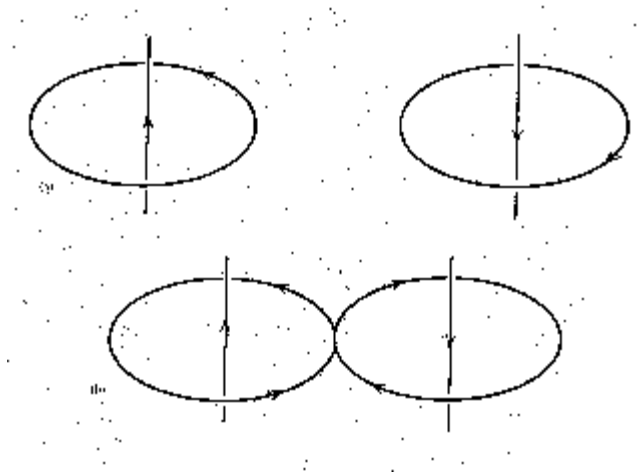}
\end{center}
\end{figure}
In this case, one expects that
\begin{equation}\label{e2C}
 E_2(R,\theta) = 2E_1 + 3k R^{-3} \cos(\theta_+ +\theta_-), 
\end{equation}
where, as before, $\theta_{\pm}$ is the angle that $\vec{P}_{\pm}$ makes
with $\vec{R}$.
So the attractive force is maximal when $\theta_+ +\theta_- = \pi$.
The dependence (\ref{e2C}) was confirmed
numerically, as before, with  $k\approx 1$.  This `maximally-attractive'
channel is referred to as channel $C$.  The effect of the discrete symmetries
$D_2$ is  $\{I,I,C,C\}$, as for channel $B$.

%%%%%%%%%%%%%%%%%%%%%%%%%%%%%%%%%%%%%%%%%%%%%%%%%%%%%%%%%%%%%%%%%%%%%%
\section{Relaxing in Channel A}

In this section, we see what happens when we begin with two solitons
far apart (in the first of the attractive channels described above), and
minimize the energy.  This was done numerically, using a conjugate-gradient
procedure.

Suppose, then, that we start in the (axisymmetric) channel $A$, and
minimize energy without breaking the axial symmetry.  Then the two solitons
approach each other along the line joining them, and the minimum is reached
when they are a nonzero distance apart.  The resulting configuration therefore
is a static solution of the field equation; it has energy $E=2.26$
(this is to be compared with $2E_1 = 2.45$), and it resembles two rings
around the $z$-axis, separated by a distance $R=1.3$.  In other words,
$W=0$ consists of the $z$-axis plus infinity, as for the 1-soliton,
while $W=\infty$ consists of two disjoint rings around it; so the linking
number $N$ does indeed equal 2.  The picture is as in Fig 1.

There is an approximate configuration analogous to (\ref{1-sol}), namely
\begin{equation}\label{1p1-sol}
 W = \frac{2(x+iy)[z(1+\beta r)-ih(r)]}
           {[z_+(1+\alpha r_+)-if(r_+)] [z_-(1+\alpha r_-)-if(r_-)]},
\end{equation} 
where $\alpha$ and $\beta$ are parameters, $f$ and $h$ are cubic polynomials,
$z_{\pm} = z \pm 0.65$ and $r_{\pm}^2 = x^2+y^2+z_{\pm}^2$.  The two-ring
structure is evident from (\ref{1p1-sol}).  Minimizing
the energy of (\ref{1p1-sol}) with respect to the ten parameters ($\alpha$,
$\beta$, and the coefficients of $f$ and $h$) gives an energy $E=2.29$,
{it ie} $1.3\%$ above that of the solution.  The corresponding configuration
$\vec{\phi}$ is very close to the actual solution.

While this is a solution, it is not the global minimum of the energy in the
$N=2$ sector; in particular, channel $B$ produces a solution with lower
energy.  So the question arises as to whether the channel $A$ minimum is
stable to (non-axisymmetric) perturbations ({\it ie} whether it is a local
minimum of the energy, as opposed to a saddle-point).  The linking behaviour
of the channel $B$ minimum is that of a single ring around a double axis (as
we shall see in the next section), as opposed to a double ring around a
single axis; there is a continuous path in configuration space from the
one configuration to
the other, but the contortions involved in this suggest that there is
an energy barrier (in other words, that the channel $A$ solution is a
local minimum).  Numerical experiments, involving random perturbations
of this solution, provide strong support for this; but more study is needed.

%%%%%%%%%%%%%%%%%%%%%%%%%%%%%%%%%%%%%%%%%%%%%%%%%%%%%%%%%%%%%%%%%%%

\section{Relaxing in Channel B}
Next, we start in channel $B$ and once again flow down the energy gradient.
As depicted in Fig 2, the two rings (where $W=\infty$) merge into one, and
then the two lines where $W=0$ merge as well.
We end up with a solution which has been described previously
\cite{gh:97}, \cite{bs:98a}, \cite{bs:98b}, and which is believed to be the
global minimum in the $N=2$ sector.  It is
axially-symmetric, and resembles a single ring; but this time
the ring winds around a double copy of the $z$-axis, and hence it has a
linking number of $N=2$.  The energy of the solution is $E = 2.00$, which
agrees with the figure given in  \cite{gh:97}.
%(the fact that the 2-soliton
%energy is 2, to an accuracy of at least two decimal places, is presumably a
%fluke).

As before, we can write down an explicit configuration which is very
close to the solution.  One such expression is
\begin{equation}\label{2-sol}
 W = \frac{(x+iy)^2}{azr-if(r)},
\end{equation} 
where $a$ is a constant and $f(r)$ is a quintic polynomial.  Minimizing
the energy with respect to the six coefficients contained in (\ref{2-sol})
gives $E=2.03$ ({\it ie} $1.5\%$ above the true minimum), for
\begin{equation} 
 a = 1.55,
 \quad f(r) = 0.23(r-1.27)(r+0.44)(r+0.16)(r^2-2.15r + 5.09).
\end{equation}
Since $f$ has only one positive root, $W=\infty$ is a ring (of radius $1.27$)
in the $xy$-plane; whereas $W=0$ is the $z$-axis, with multiplicity two.
The components of $\vec{\phi}$ derived from (\ref{2-sol}) are very close to
those of the actual solution.
%Note that the channel-$B$ discrete symmetries are evident from (\ref{2-sol}):
%the four elements of $D_2$ induce the four maps $\{I,I,-C,-C\}$ on $W$.

%%%%%%%%%%%%%%%%%%%%%%%%%%%%%%%%%%%%%%%%%%%%%%%%%%%%%%%%%%%%%%%%%%%
\section{Relaxing in Channel C}

If one begins with the configuration depicted in Fig 3(a) and moves in the
direction of the energy gradient, the two solitons approach each other.
If the two $W=\infty$ loops touch, one has a figure-eight curve, with the
$W=0$ lines linking through it in opposite directions: Fig 3(b).  This
configuration is certainly not stable: preliminary numerical work indicates
that the two `halves' of the configuration rotate by $\pi/2$ (in opposite
directions) about the axis joining them.  So the figure-eight untwists to
become a simple loop, and the two $W=0$ curves end up pointing in the same
direction, exactly as in Fig 2(b) and (c).  Hence the minimum in channel $C$
is the same as that in channel $B$.  Between this mimimum and the channel-$A$
one, there should be saddle-point solutions; but what these look like is not
yet clear.

%%%%%%%%%%%%%%%%%%%%%%%%%%%%%%%%%%%%%%%%%%%%%%%%%%%%%%%%%%%%%%%%%%%%
\section{Concluding Remarks}

There has already been some study of two-soliton dynamics, using a
``direct'' numerical approach (see, for example, \cite{hs:99}); this is
computationally very intensive.  The results reported in this Letter
could be viewed as the first step towards a somewhat different approach,
namely that of constructing a collective-coordinate manifold for the
two-soliton system.  The analogous structure for the Skyrme model has been
investigated in some detail \cite{m:88}, \cite{am:93}; in particular, it
has the advantage that one can introduce quantum corrections by quantizing
the dynamics on the collective-coordinate manifold \cite{lms:95}.
Since each Hopf soliton depends on six parameters, the two-soliton
manifold  $M_2$ should have dimension (at least) twelve; each point
of $M_2$ corresponds to a relevant $N=2$ configuration, and the expressions
(\ref{1-sol}) and (\ref{2-sol}) are examples of such configurations.

But clearly much more work remains to be done towards understanding the energy
functional on the $N=2$ configuration space.  The suggestion of this
Letter is that the global minimum (which is, of course, degenerate: it
depends on six moduli) is as in Fig 2(c); there is a local minimum as in
Fig 1; and between the two are saddle-point solutions which may be related to
the figure-eight configuration Fig 3(b).

%%%%%%%%%%%%%%%%%%%%%%%%%%%%%%%%%%%%%%%%%%%%%%%%%%%%%%%%%%%%%%%%%%%
\bibliographystyle{plain} \bibliography{paper_ref}

\begin{thebibliography}{10}

\bibitem{fn:99}
L~Faddeev, A~J~Niemi, Phys Rev Lett 82 (1999) 1624; hep-th/9807069

\bibitem{s:99}
S~V~Shabanov, Phys Lett B 458 (1999) 322.

\bibitem{clp:99}
Y~M~Cho, H~W~Lee, D~G~Pak, Effective Theory of QCD, hep-th/9905215

\bibitem{f:75}
L~Faddeev, Quantisation of Solitons [Preprint IAS
           Print-75-QS70, Princeton]; Lett Math Phys 1 (1976) 289.

\bibitem{v:78}
H~J~deVega, Phys Rev D 18 (1978) 2945.

\bibitem{fn:97}
L~Faddeev, A~J~Niemi, Nature 387 (1997) 58.

\bibitem{gh:97}
J~Gladikowski, M~Hellmund, Phys~Rev~D 56 (1997) 5194

\bibitem{bs:97}
R~A~Battye, P~M~Sutcliffe, Solitonic Strings and Knots.
           To appear in the CRM Series in Mathematical Physics
           (Springer-Verlag).

\bibitem{bs:98a}
R~A~Battye, P~M~Sutcliffe, Phys Rev Lett 81 (1998) 4798; hep-th/9808129

\bibitem{bs:98b}
R~A~Battye, P~M~Sutcliffe, Proc R Soc Lond A 455 (1999) 4305; hep-th/9811077

\bibitem{hs:99}
J~Hietarinta, P~Salo P, Phys Lett B 451 (1999) 60; hep-th/9811053

\bibitem{w:99}
R~S~Ward, Nonlinearity 12 (1999) 1; hep-th/9811176

\bibitem{g:98}
T~R~Govindarajan, Mod Phys Lett A 13 (1998) 3179; hep-th/9811171

\bibitem{n:99}
A~J~Niemi, Knots in Interaction, hep-th/9902140

\bibitem{m:88}
N~S~Manton, Phys Rev Lett 60 (1988) 1916.

\bibitem{am:93}
M~F~Atiyah, N~S~Manton, Commun Math Phys 152 (1993) 391.

\bibitem{lms:95}
R~A~Leese, N~S~Manton, B~J~Schroers, Nucl Phys B 442 (1995) 228.

\end{thebibliography}

\end{document}